\documentstyle[12pt]{article}

\begin{document}
{\it University of Shizuoka}

\hspace*{10.5cm} {\bf US-00-04R}\\[-.3in]

\hspace*{10.5cm} {\bf July 2000}\\[-.3in]

\hspace*{10.5cm} {\bf hep-ph/0006084}\\[.3in]

\vspace*{.4in}

\begin{center}

{\Large\bf  Democratic Universal Seesaw Model \\[.1in]

with Three Light Sterile Neutrinos} \\[.3in]

{\bf Yoshio Koide}\footnote{
E-mail: koide@u-shizuoka-ken.ac.jp} {\bf and
Ambar Ghosal}\footnote{
E-mail: gp1195@mail.a.u-shizuoka-ken.ac.jp} \\

Department of Physics, University of Shizuoka \\ 
52-1 Yada, Shizuoka 422-8526, Japan \\[.1in]

\vspace{.3in}

{\large\bf Abstract}\\[.1in]

\end{center}

\begin{quotation}
Based on the ``democratic" universal seesaw model, 
where mass matrices $M_f$ of quarks and leptons $f_i$ 
($f=u,d,\nu,e$; $i=1,2,3$) are 
given by a seesaw form $M_f\simeq-m_LM_F^{-1}m_R$, and 
$m_L$ and $m_R$ are universal for all the fermion sectors, 
and the mass matrices $M_F$ of hypothetical 
heavy fermions $F_i$ have a democratic structure, a possible 
neutrino mass matrix is investigated.  
In the model, there are three sterile neutrinos $\nu_{iR}$
which mix with the active neutrinos $\nu_{iL}$ with 
$\theta \sim 10^{-2}$ and which are harmless for 
constraint from the big bang nucleosynthesis.
The atmospheric, solar and the LSND neutrino data are 
explained by the mixings $\nu_{{\mu}L}
\leftrightarrow\nu_{{\tau}L}$, $\nu_{eL}\leftrightarrow
\nu_{eR}$ and $\nu_{eL}\leftrightarrow\nu_{{\mu}L}$, respectively.  
The model predicts that 
${\Delta}m^2_{solar}/{\Delta}m^2_{atm}{\simeq}
(R^2-1) m_e/{\sqrt{m_{\mu}m_{\tau}}}$ [$R=m(\nu_{iR})/m(\nu_{iL})$
($i=1,2,3$)] with $\sin^{2}2\theta_{atm}\simeq1$ 
and
${\Delta} m_{LSND}^2/{\Delta} m^2_{atm}{\simeq}(1/4)
{\sqrt{m_{\mu}/m_e}}$ 
with  $\sin^{2}2\theta_{LSND}{\simeq}4m_e/m_{\mu}$.
\end{quotation}

\vfill
PACS numbers: {14.60.Pq, 14.60.St, 12.60.-i}
\newpage
\section{Introduction}

In order to seek for a clue to the unified understanding of quarks and 
leptons, many attempts to give a unified description of the quark and 
lepton mass matrices have been proposed.  The universal seesaw mass 
matrix model \cite{USM} is one of the promising attempts to view the 
unified description, 
where the mass matrices $M_f$ for the conventional quarks and leptons 
$f_i$ ($f=u,d,\nu,e$; $i=1,2,3$) are given by 
$$
(\overline{f}_L\ \overline{F}_L)
\left(\begin{array}{cc}
0 & m_L \\
m_R & M_F 
\end{array}\right)
\left(\begin{array}{c}
f_R \\
F_R
\end{array}\right),
\eqno(1.1)
$$
and $m_L$ and $m_R$ are universal for all fermion sectors $f$.  
For $O(M_F){\gg}O(m_R){\gg}O(m_L)$, the mass matrix 
(1.1) leads to the well-known seesaw expression
$$
M_f\simeq-m_LM_F^{-1}m_R.
\eqno(1.2)
$$

As a specific version of such universal seesaw model, Fusaoka and 
one of the authors (Y.K.) have proposed a so-called ``democratic" 
seesaw model \cite{KF-zp}: The heavy fermion matrices $M_F$ have a simple 
structure [(unit matrix)+(democratic matrix)], i.e.,
$$
M_F=m_0\lambda_f({\bf 1}+3b_fX),
\eqno(1.3)
$$
$$
{\bf 1}=
\left(\begin{array}{ccc}
1 & 0 & 0 \\
0 & 1 & 0 \\
0 & 0 & 1 
\end{array}\right),\ \ \ 
X=
{\frac{1}{3}}
\left(\begin{array}{ccc}
1 & 1 & 1 \\
1 & 1 & 1 \\
1 & 1 & 1 
\end{array}\right),
\eqno(1.4)
$$
on the basis on which the matrices $m_L$ and $m_R$ are diagonal:
$$
m_L={\frac{1}{\kappa}}m_R=m_{0}Z=
m_0
\left(\begin{array}{ccc}
z_1 & 0 & 0 \\
0 & z_2 & 0 \\
0 & 0 & z_3 
\end{array}\right),
\eqno(1.5)
$$
where the parameters $z_1$, $z_2$ and $z_3$ are normalized as 
$z_1^2+z_2^2+z_3^2=1$, 
and $m_0$ is of the order of the electroweak symmetry breaking scale, 
i.e., $m_0\sim10^2$ GeV.  
Since the parameter $b_f$ in the charged lepton 
sector is taken as $b_e=0$, the parameters $z_i$ are fixed as 
$$
{\frac{z_1}
{{\sqrt{m_e}}}}
=
{\frac{z_2}
{{\sqrt{m_{\mu}}}}}
=
{\frac{z_3}
{{\sqrt{m_{\tau}}}}}
=
{\frac{1}
{\sqrt{m_{\tau}+m_{\mu}+m_e}}}.
\eqno(1.6)
$$
For the up-type quark sector, the parameter $b_f$ is taken as $b_u=-1/3$,  
which leads to det$M_U=0$, and the seesaw mechanism does not 
work for one of the three families, and hence we obtain 
the mass $m_t\simeq m_0/{\sqrt{3}}$ without the seesaw suppression 
factor ${\kappa}/{\lambda}_u$ 
(we identify it as the top quark mass).  
Furthermore, we also obtain a relation $m_u/m_c\simeq3m_e/m_{\mu}$, 
which is in good agreement with the observed values.  
Moreover, when we take $b_d\simeq{-1}$ 
($b_d=-e^{i{\beta}_d}$ with ${\beta}_d=18^{\circ}$) for the down-type
quark sector, we can obtain reasonable quark mass ratios and 
the Cabbibo-Kobayashi-Maskawa \cite{CKM} (CKM) matrix.

The neutrino mass matrix in the universal seesaw mass matrix model is given 
as follows:
$$
\left(\begin{array}{cccc}
\overline{\nu}_L & \overline{\nu}^c_R & \overline{N}_L & 
\overline{N}^c_R
\end{array}\right)
\left(\begin{array}{cccc}
0 & 0 & 0 & m_L \\ 
0 & 0 & m_R^T & 0 \\
0 & m_R & M_{NL} & M_D \\
m_L^T & 0 & M_D^T & M_{NR} 
\end{array}\right)
\left(\begin{array}{c}
\nu_L^c \\
\nu_R \\
N_L^c \\
N_R 
\end{array}\right),
\eqno(1.7)
$$
where $\psi_L^c\equiv(\psi_L)^c=C\psi_L^T$.  
[We consider a SO(10)$_L\times$SO(10)$_R$  model \cite{SO10}, 
where fermions $(f_L+F_R^c)$ and $(f_R+F_L^c)$ 
are assigned to (16,1) and (1,16) under SO(10)$_L\times$SO(10)$_R$,
respectively. 
Hereafter, we will denote the Majorana 
mass matrices $M_{NL}$ and $M_{NR}$ of the neutral heavy leptons 
$N_L$ and $N_R$ as $M_R=M_{NL}$ and $M_L=M_{NR}$, respectively.] 

For $O(m_L){\ll}O(m_R){\ll}O(M_D),O(M_L),O(M_R)$, we obtain the 
following $6\times 6$ seesaw mass matrix for $(\nu_L^c,\nu_R)$
$$
M^{(6\times6)}{\simeq}-
\left(\begin{array}{cc}
0 & m_L \\
m_R^T & 0 
\end{array}\right)
\left(\begin{array}{cc}
M_R & M_D \\
M_D^T & M_L 
\end{array}\right)^{-1}
\left(\begin{array}{cc}
0 & m_R \\
m_L^T & 0
\end{array}\right),
\eqno(1.8)
$$
which leads to the 3$\times$3 seesaw matrices for 
$\nu_L$ and $\nu_R$ 
$$
M(\nu_L)\simeq-m_LM_L^{-1}m_L^T,
\eqno(1.9)
$$
$$
M(\nu_R)\simeq-m_RM_R^{-1}m_R^T.
\eqno(1.10)
$$
The scenario corresponding to 
$O(m_LM_L^{-1}m_L^T){\ll}O(m_RM_R^{-1}m_R^T)$ has 
already been 
investigated by one of the authors (Y.K.) \cite{Koide-nu}.  
He has concluded 
that although either the atmospheric \cite{atm} or solar 
\cite{solar} neutrino data can be 
explained by the mixings $\nu_{\mu}\leftrightarrow\nu_{\tau}$ or 
$\nu_e\leftrightarrow\nu_{\mu}$, however, simultaneous explanation of 
the both data cannot be obtained in this model.  

In the present paper, we consider another possibility 
$O(m_LM_L^{-1}m_L^T){\sim}O(m_RM_R^{-1}m_R^T)$.  
In this case, mixings between 
$\nu_{iL}$ and $\nu_{iR}$ are induced.  
The solar neutrino data \cite{solar} are 
understood from a small mixing between $\nu_{eL}$ and $\nu_{eR}$.  
The atmospheric \cite{atm} and the LSND \cite{LSND} neutrino data 
are explained by the 
mixings $\nu_{{\mu}L}\leftrightarrow\nu_{{\tau}L}$ and 
$\nu_{eL}\leftrightarrow\nu_{{\mu}L}$, respectively.  
The vantage point of the democratic seesaw model \cite{KF-zp} is that 
parameters $z_i$ in the mass matrices $m_L$ and $m_R$ are given 
in terms of the charged lepton masses  and thereby the mass spectrum 
and mixings of $\nu_{iL}$ and $\nu_{iR}$ can also be predicted 
in terms of the charged lepton masses.  

\section{Parameter $b_{\nu}$}

In the present paper, for simplicity, we assume that all the 
neutral heavy fermion mass matrices $M_D$, $M_L$ and $M_R$ have 
the same flavor structure
$$
{\frac{1}{\lambda_D}}M_D=
{\frac{1}{\lambda_L}}M_L=
{\frac{1}{\lambda_R}}M_R=
m_0({\bf 1}+3b_{\nu}X),
\eqno(2.1)
$$
and we will investigate only the case $b_{\nu}=-1/2$.

The excuse for considering only the case $b_{\nu}=-1/2$ is as follows.  
The choices of $b_f$ ($b_e=0, b_u=-1/3, b_d\simeq-1$) have given the 
successful description of the quark masses and mixings in terms of the 
charged lepton masses.  When, instead of the expression (1.3), 
we denote $M_F$ as 
$$
M_F= m_0 \lambda_f \sqrt{1+2 b_f +3 b_f^2}  (\cos\phi_{f}E-\sin\phi_{f}S),
\eqno(2.2)
$$
$$
E=
{\frac{1}{\sqrt{3}}}
\left(\begin{array}{ccc}
1 & 0 & 0 \\
0 & 1 & 0 \\
0 & 0 & 1 
\end{array}\right),\ \ \ 
S=
{\frac{1}{{\sqrt{6}}}}
\left(\begin{array}{ccc}
0 & 1 & 1 \\
1 & 0 & 1 \\
1 & 1 & 0 
\end{array}\right),
\eqno(2.3)
$$
where $E$ and $S$ have been normalized as ${\rm Tr}E^2= {\rm Tr}S^2=1$
and $\tan\phi_f = -\sqrt{2} b_f/(1+b_f)$,
the cases $b_e=0$, $b_u=-1/3$ and 
$b_d=-1$ correspond to $(\cos\phi_f, \sin\phi_f)=(1,0)$, 
$({\sqrt{2/3}},{\sqrt{1/3}})$ and $(0,1)$, respectively.  
Considering an empirical relation $\phi_d=\pi/2-\phi_e$ for 
$(\cos\phi_e, \sin\phi_e)=(1,0)$ 
and $(\cos\phi_d, \sin\phi_d)=(0,1)$, 
we consider that the value of $b_{\nu}$ 
is also given by $\phi_{\nu}=\pi/2-\phi_u$ for 
$(\cos\phi_u, \sin\phi_u)=({\sqrt{2/3}},{\sqrt{1/3}})$, 
i.e., we assume
$$
(\cos\phi_{\nu}, \sin\phi_{\nu})=({\sqrt{1/3}}, {\sqrt{2/3}}),
\eqno(2.4)
$$
which corresponds to the case $b_{\nu}=-1/2$.  

Besides, from the phenomenological point of view, the case $b_{\nu}=-1/2$ 
is also interesting.  The inverse matrix of the $M_L$ with $b_{\nu}=-1/2$
$$
M_L=m_0{\lambda}_L({\bf 1}-{\frac{1}{2}}\cdot3X)=
{\frac{1}{2}}m_0{\lambda}_L
\left(\begin{array}{ccc}
1 & -1 & -1 \\
-1 & 1 & -1 \\
-1 & -1 & 1 
\end{array}\right),
\eqno(2.5)
$$
is given by
$$
M_L^{-1}=-{\frac{1}
{m_0{\lambda}_L}}
\left(\begin{array}{ccc}
0 & 1 & 1 \\
1 & 0 & 1 \\
1 & 1 & 0 
\end{array}\right),
\eqno(2.6)
$$
so that the seesaw matrix $M_{\nu}{\simeq} -m_L M_L^{-1} m_L^T$ is 
expressed as 
$$
M_{\nu}\simeq
m_0{\frac{1}{\lambda_L}}
\left(\begin{array}{ccc}
0 & z_1z_2 & z_1z_3 \\
z_1z_2 & 0 & z_2z_3 \\
z_1z_3 & z_2z_3 & 0 
\end{array}\right).
\eqno(2.7)
$$
The form (2.7) is just a Zee-type mass matrix \cite{Zee}, which has recently 
been revived \cite{recentZee} as a promising neutrino mass matrix 
form.

\section{Mass spectrum and mixing}

For the specific form (2.1) with $b_{\nu}=-1/2$, the 6$\times$6 
seesaw matrix $M^{(6\times6)}$ given by Eq.~(1.8) becomes
$$
M^{(6\times6)}\simeq
-m_0
\left(\begin{array}{cc}
0 & Z \\
{\kappa} Z & 0 
\end{array}\right)
\left(\begin{array}{cc}
\lambda_{R}Y & \lambda_{D}Y \\
\lambda_{D}Y & \lambda_{L}Y 
\end{array}\right)^{-1}
\left(\begin{array}{cc}
0 & {\kappa} Z \\
Z & 0 
\end{array}\right)
$$
$$
=-m_0
{\frac{1}
{\lambda_R\lambda_L-\lambda_D^2}}
\left(\begin{array}{cc}
\lambda_{R}ZY^{-1}Z & -\kappa\lambda_{D}ZY^{-1}Z \\
-\kappa\lambda_{D}ZY^{-1}Z & \kappa^2\lambda_{L}ZY^{-1}Z 
\end{array}\right),
\eqno(3.1)
$$
where
$$
Y={\bf 1}+3b_{\nu}X,\ \ \ Y^{-1}={\bf 1}+3a_{\nu}X,
\eqno(3.2)
$$
$$
a_{\nu}=-b_{\nu}/(1+3b_{\nu}).
\eqno(3.3)
$$
Therefore, the matrix $M^{(6\times6)}$ is diagonalized by the 6$\times$6 unitary 
matrix $U^{(6\times6)}$
$$
U^{(6\times6)}=
\left(\begin{array}{cc}
\cos\theta\cdot{U} & -\sin\theta\cdot{U} \\
\sin\theta\cdot{U} & \cos\theta\cdot{U} 
\end{array}\right),
\eqno(3.4)
$$
as
$$
U^{(6\times6)\dagger}M^{(6\times6)}U^{(6\times6)}=
{\rm diag}(m_{\nu_{1L}}, m_{\nu_{2L}}, m_{\nu_{3L}},
m_{\nu_{1R}}, m_{\nu_{2R}}, m_{\nu_{3R}})
$$
$$
= m_0{\rm diag}(\xi_L\rho_1, \xi_L\rho_2, \xi_L\rho_3,
\xi_R\rho_1, \xi_R\rho_2, \xi_R\rho_3),
\eqno(3.5)
$$
where
$$
U^{\dagger}ZY^{-1}ZU=
{\rm diag}(\rho_1, \rho_2, \rho_3),
\eqno(3.6)
$$
$$
\left(\begin{array}{cc}
\cos\theta & \sin\theta \\
-\sin\theta & \cos\theta
\end{array}\right)
\left(\begin{array}{cc}
\lambda_R & -\kappa\lambda_D \\
-\kappa\lambda_D & \kappa^2\lambda_L
\end{array}\right)
\left(\begin{array}{cc}
\cos\theta & -\sin\theta \\
\sin\theta & \cos\theta
\end{array}\right)
=
\left(\begin{array}{cc}
\lambda'_L & 0 \\
0 & \lambda'_R
\end{array}\right),
\eqno(3.7)
$$
$$
\xi_L={\frac{\lambda'_L}
{\lambda_R\lambda_L-\lambda^2_D}} \ ,\ \ \ 
\xi_R={\frac{\lambda'_R}
{\lambda_R\lambda_L-\lambda^2_D}} \ ,
\eqno(3.8)
$$
$$
\left(\begin{array}{c}
\lambda'_L \\
\lambda'_R
\end{array}\right)
={\frac{1}{2}}
(\lambda_R+\kappa^2\lambda_L)
\mp {\frac{1}{2}}
(\lambda_R-\kappa^2\lambda_L)
{\sqrt{1+\tan^{2}2\theta}} \ .
\eqno(3.9)
$$
The mixing angle $\theta$ between $\nu_{iL}$ and $\nu_{iR}$ is given by 
$$
\tan2\theta=
{\frac{2\kappa\lambda_D}
{\lambda_R-\kappa^2\lambda_L}}.
\eqno(3.10)
$$
The light neutrino masses $m(\nu_{iL})$ and $m(\nu_{iR})$ are given by 
$$
m(\nu_{iL})=m_0\xi_{L}\rho_i \ ,\ \ \ 
m(\nu_{iR})=m_0\xi_{R}\rho_i \ .
\eqno(3.11)
$$
For the case of $b_{\nu}=-1/2$, the eigenvalues $\rho_i$ of 
the matrix $ZY^{-1}Z$ are given by 
$$
\rho_1\simeq-2z_1^2, \ \ 
\rho_2\simeq-\left(z_2+{\frac{z_1^2}{2z_2}}-z_1^2\right), \ \ 
\rho_3{\simeq}z_2+{\frac{z_1^2}{2z_2}}+z_1^2,
\eqno(3.12)
$$
so that
$$
\rho_3^2-\rho_2^2\simeq4z_2z_1^2, \ \ 
\rho_2^2-\rho_1^2{\simeq}z_2^2.
\eqno(3.13)
$$
The $3\times3$ mixing matrix $U$ for the case $b_{\nu}=-1/2$ is given by
$$
U\simeq
\left(\begin{array}{ccc}
-1 & -{\frac{1}{{\sqrt{2}}}}{\frac{z_1}{z_2}}(1-z_2) & 
{\frac{1}{{\sqrt{2}}}}{\frac{z_1}{z_2}}(1+z_2) \\
{\frac{z_1}{z_2}} & -{\frac{1}{{\sqrt{2}}}} & {\frac{1}{{\sqrt{2}}}} \\
z_1 & {\frac{1}{{\sqrt{2}}}} & {\frac{1}{{\sqrt{2}}}}
\end{array}\right).
\eqno(3.14) 
$$

\section{Explanations of the neutrino data}

The atmospheric \cite{atm} and solar \cite{solar} neutrino data 
are explained by the mixings $\nu_{{\mu}L}\leftrightarrow\nu_{{\tau}L}$ 
and $\nu_{eL}\leftrightarrow\nu_{eR}$, respectively.  
As seen in the mixing matrix
(3.14), the neutrinos $\nu_{{\mu}L}$ and $\nu_{{\tau}L}$ are 
maximally mixed.
On the other hand, the mixing between 
$\nu_{eL}$ and $\nu_{eR}$ is given by Eq.~(3.10).  Since the solar neutrino 
data disfavor \cite{Hata} sterile neutrino with a large mixing angle, 
we take the small mixing angle solution in the Mikheyev-Smirnov-Wolfenstein 
(MSW) mechanism \cite{MSW}, 
$$
{\Delta}m_{solar}^2\simeq 4.0 \times10^{-6} \ {\rm eV}^2,\ \ 
\sin^{2}2\theta_{solar} \simeq 6.9 \times10^{-3}.
\eqno(4.1)
$$
Here, the values in Eq.~(4.1) have been quoted from the recent analysis 
for $\nu_{e}\rightarrow\nu_{s}$ by Bahcall, Krastev and Smirnov 
\cite{Bahcall}.  The value $\sin^{2}2\theta_{solar}\simeq7\times10^{-3}$ 
can be fitted by adjusting the parameters $\lambda_L$, 
$\lambda_R/\kappa^2$ and $\lambda_D/\kappa$ in Eq.~(3.10).

As seen from Eqs.~(3.5) and (3.13), the ratio of 
${\Delta}m^2_{solar}=(m_{{\nu}_{1R}})^2-(m_{{\nu}_{1L}})^2$ to 
${\Delta}m^2_{atm}=(m_{{\nu}_{3L}})^2-(m_{{\nu}_{2L}})^2$ is given by
$$
{\frac{{\Delta}m^2_{solar}}{{\Delta}m^2_{atm}}}\simeq
\frac{\lambda_R^{\prime 2}- \lambda_L^{\prime 2}}{\lambda_L^{\prime 2}}
\frac{4z_{1}^{4}}{4z_{2}z_{1}^{2}} \simeq
(R^2 -1) \frac{m_e}{\sqrt{m_{\mu}m_{\tau}}} = (R^2-1)\times 1.15
\times 10^{-3},
\eqno(4.2)
$$
where
$$
R=
{\frac{\lambda'_R}{\lambda'_L}}=
{\frac{\xi_R}{\xi_L}}=
{\frac{m(\nu_{iR})}{m(\nu_{iL})}}.
\eqno(4.3)
$$
The recent best fit value $\Delta m^2_{atm}=3.2\times 10^{-3}$
eV$^2$ \cite{atm-nu2000} gives the ratio
$$
{\frac{{\Delta}m^2_{solar}}{{\Delta}m^2_{atm}}}\simeq
{\frac{4.0 \times10^{-6}{\rm eV}^2}{3.2 \times10^{-3}{\rm eV}^2}}
\simeq1.3\times10^{-3}.
\eqno(4.4)
$$
By comparing Eqs.~(4.2) and (4.4), we obtain $R\simeq 1.4$.
Note that the observed value (4.4) is in good agreement
with the value $m_e/\sqrt{m_\mu m_\tau}$, so that we are tempted
to consider a model with $R \simeq 0$. 
However, the sign of $\Delta m^2_{solar}$
in the small mixing angle MSW solution must be positive, so 
that we cannot consider the case $R\simeq 0$.
In the present model, $R$ is only a phenomenological parameter
with the constraint $R >1$.

The LSND data \cite{LSND} is explained by the mixing 
$\nu_{eL}\leftrightarrow\nu_{eR}$. 
The mass-squared difference 
${\Delta}m^2_{LSND}= m_{{\nu}_{2L}}^2- m_{{\nu}_{1L}}^2$ 
and the $\nu_{eL}\leftrightarrow\nu_{eR}$ mixing angle 
are given by the ratio
$$
{\frac{{\Delta}m^2_{LSND}}{{\Delta}m^2_{atm}}}\simeq
{\frac{z_2}{4z_1}}\simeq{\frac{1}{4}}
{\sqrt{\frac{m_{\mu}}{m_e}}}=
2.2\times10^2,
\eqno(4.5)
$$
and 
$$
\sin^{2}2\theta_{LSND}\simeq 4U_{e1}^2 U_{\mu1}^2 
\simeq 4\left(
{\frac{z_1}{z_2}}\right)^2\simeq
4{\frac{m_e}{m_{\mu}}}\simeq
0.019 \ ,
\eqno(4.6)
$$
respectively.
The best fit value $\Delta  m^2_{atm} \simeq 3.2 \times 10^{-3}$ 
${\rm eV}^2$ give a prediction $\Delta m^2_{LSND} \simeq 0.70$
${\rm eV}^2$.
However, the region $\Delta m^2_{LSND} \geq 0.34$ ${\rm eV}^2$
in the LSND favored region at $\sin^2 2\theta=0.02$ has been 
excluded by the recent KARMEN2 experiment \cite{Karmen}.
Therefore, only when we take the value $\Delta m^2_{LSND} 
\simeq 0.33$ ${\rm eV}^2$, we can obtain the prediction
$\Delta m^2_{atm} \simeq 1.5 \times 10^{-3}$ ${\rm eV}^2$
which is barely inside the 90\% C.L. allowed region
($ 1.5 \times 10^{-3} \ {\rm eV}^2 \leq \Delta m^2_{atm} 
\leq 5 \times 10^{-3} \ {\rm eV}^2$) in the recent 
Super-Kamiokande atmospheric neutrino data \cite{atm-nu2000}.
Hereafter, we will adopt this pinpoint solution:
$$
\Delta m^2_{LSND} \simeq 0.33\  {\rm eV}^2\ , \ \ \ 
\Delta m^2_{atm} \simeq 1.5 \times 10^{-3}\ {\rm eV}^2 \ .
\eqno(4.7)
$$
Then, the parameter $R$ is fixed as
$$
R\simeq 1.8 \ ,
\eqno(4.8)
$$
from Eq.~(4.2), and the neutrino masses are predicted as follows:
$$
m(\nu_{3L}) \simeq m(\nu_{2L}) \simeq  0.57 \ {\rm eV} \ , \ \ 
m(\nu_{1L})\simeq 1.3  \times 10^{-3} \ {\rm eV} \ ,
\eqno(4.9)
$$
$$
m(\nu_{3R}) \simeq m(\nu_{2R}) \simeq  1.05 \ {\rm eV} \ , \ \ 
m(\nu_{1R}) \simeq 2.5\times 10^{-3} \ {\rm eV} \ ,
\eqno(4.10)
$$
where we have used the relation 
$m(\nu_{2L}) \simeq \sqrt{\Delta m^2_{LSND}}$. 

In the present scenario, there are three light sterile neutrinos 
$\nu_{iR}\ (i=1,2,3)$.  However, those neutrinos do not spoil the big bang 
nucleosynthesis (BBN) scenario, which puts the following constraint \cite{BBN} 
for a mixing between the active neutrino $\nu_{\alpha}\ (\alpha=e,\mu,\tau)$ 
and a sterile neutrino $\nu_s$, 
$$
(\sin^{2}2\theta_{{\alpha}s})^2{\Delta}m^2_{{\alpha}s}<
3.6\times10^{-4}\ {\rm eV}^2.
\eqno(4.11)
$$
The value of $(\sin^{2}2\theta)^2{\Delta}m^2$ 
in our model is less than $10^{-4}\ {\rm eV}^2$, because the mixing angle 
$\theta$ in the present model is sufficiently small, i.e., 
$(\sin^{2}2\theta)^2=(6.9 \times10^{-3})^2 =4.8 \times10^{-5}$.

However, we have another severe constraint on the neutrino 
masses from the cosmic structure formation in a low-matter-density universe 
\cite{Fukugita}
$$
N_{\nu}m_{\nu}<1.8\ {\rm eV}\ \ (1.5\ {\rm eV}),
\eqno(4.12)
$$
for flat universe (for open universes), where $N_{\nu}$ is the number of 
almost degenerate neutrinos with the highest mass.  
The present model 
gives $N_{\nu}m_{\nu} \simeq 3.2$ ${\rm eV}$, so that 
the model dose not satisfies the constraint (4.12).
We will go optimistically for this problem.

The mixing between $\nu_{eL}$ and $\nu_{{\tau}L}$ is given by 
$$
U_{e3}\simeq{\frac{1}{\sqrt{2}}}{\frac{z_1}{z_2}}
(1+z_2)\simeq{\sqrt{\frac{m_e}{2m_{\mu}}}}
\left(1+{\sqrt{\frac{m_{\mu}}{m_{\tau}}}}
\right)\simeq0.061,
\eqno(4.13 )
$$
which safely satisfies the constraint $|U_{e3}|\leq(0.22-0.14)$ obtained from 
the CHOOZ reactor neutrino experiment \cite{CHOOZ}.

\section{Conclusion and discussion}

In conclusion, we have investigated a neutrino mass matrix in the framework of 
the ``democratic" universal seesaw model.  Although the model has three light 
sterile neutrinos $\nu_{iR}$ $(i=1,2,3)$, they do not spoil the BBN scenario, 
because the mixing angle $\theta$ between the active and sterile neutrinos is 
taken as $\sin^{2}2\theta\simeq7\times10^{-3}$.  The atmospheric, solar and 
LSND neutrino data are explained by the mixings $\nu_{{\mu}L}\leftrightarrow
\nu_{{\tau}L}$, $\nu_{eL}\leftrightarrow\nu_{eR}$ and $\nu_{eL}\leftrightarrow
\nu_{{\mu}L}$, respectively.  The model with the parameter $b_{\nu}=-1/2$ 
gives the predictions in terms of the charged lepton masses,
$$
{\frac{{\Delta}m^2_{solar}}{{\Delta}m^2_{atm}}}\simeq
(R^2 -1){\frac{m_e}{{\sqrt{m_{\mu}m_{\tau}}}}}\ , \ \ \ 
{\frac{{\Delta}m^2_{LSND}}{{\Delta}m^2_{atm}}}\simeq
{\frac{1}{4}}
{\sqrt{\frac{m_{\mu}}{m_e}}}\ ,
\eqno(5.1)
$$
$$
\sin^{2}2\theta_{atm}\simeq1\ ,\ \ \ \sin^{2}2\theta_{LSND}\simeq4
{\frac{m_e}{m_{\mu}}}\ ,
\eqno(5.2)
$$
where $R=m(\nu_{iR})/m(\nu_{iL})$.
In the present model, the prediction 
${\Delta}m^2_{solar}/{\Delta}m^2_{atm}$ includes a free parameter $R$.
Only a parameter independent prediction is 
${\Delta}m^2_{LSND}/{\Delta}m^2_{atm}$ together with 
$\sin^{2}2\theta_{LSND}\simeq 4m_e/m_{\mu}$.
Since the most part of the allowed region of the $\nu_e$-$\nu_\mu$
oscillation in the LSND data is ruled out by the KARMEN2 data
\cite{Karmen} (but a narrow region still remains), the predictability 
of the present model is somewhat faded from the point of view of
the neutrino phenomenology.
However, the motivation of the present paper is not to give the
explanation of the LSND data, but to seek for a possible unification
model of the quark and lepton mass matrices.
The presence of the light right-handed neutrinos $\nu_{iR}$ will
offer fruitful new physics to the near future neutrino
experiments.

In the present scenario, the following intermediate energy scales have 
been considered: The neutral leptons $N_L$ and $N_R$ acquire large Majorana
 masses $M_R$ and $M_L$ at ${\mu}=\Lambda_{NL}=m_0\lambda_R$ and 
$\mu=\Lambda_{NR}=m_0\lambda_L$, respectively.  The fermions $N$ and $F$ 
$(F=U,D,E)$ acquire large Dirac masses $M_D$ and $M_F$ at $\mu=\Lambda_D
=m_0\lambda_D$ and $\mu=\Lambda_F=m_0\lambda_F$, respectively.  The gauge 
symmetries SU(2)$_R$ and SU(2)$_L$ are broken at $\mu=\Lambda_R=m_{0}\kappa$ 
and $\mu=\Lambda_L=m_0$, respectively. 
For $\tan^2 2\theta \ll 1$, form Eq.~(3.9), we obtain the approximate
relations
$$
\lambda'_L \simeq \kappa^2 \lambda_L \ , \ \ \ 
\lambda'_R \simeq \lambda_R \ ,
\eqno(5.3)
$$
so that
$$
R = \frac{\lambda'_R}{\lambda'_L} \simeq  
\frac{\lambda_R}{\kappa^2 \lambda_L} \ .
\eqno(5.4)
$$
The numerical result $R=O(1)$ means $\lambda_R/\lambda_L \sim \kappa^2$,
i.e.,
$$
\frac{\Lambda_{NL}}{\Lambda_{NR}} \sim \left( 
\frac{\Lambda_R}{\Lambda_L} \right)^2 \ .
\eqno(5.5)
$$
Since
$$
m(\nu_{2L}) = \xi_L \rho_2 m_0 \simeq  \frac{\kappa^2 \lambda_L}{
\lambda_R \lambda_L - \lambda_D^2} \rho_2 m_0 \simeq
\frac{1}{\lambda_R/\kappa^2} \sqrt{\frac{m_\mu}{m_\tau}} m_0 \ ,
\eqno(5.6)
$$
we estimate
$$
\frac{\lambda_R}{\kappa^2} \simeq \sqrt{\frac{m_\mu}{m_\tau}}
\frac{m_0}{m(\nu_{2L})} \sim 10^{11}\ , 
\eqno(5.7)
$$
where we have used $m_0 \sim 10^2$ GeV, so that we obtain
$\Lambda_{NL} \sim \kappa^2 \times 10^{13}$ GeV.
If we consider that $\Lambda_{NL}$ must be smaller than the
Planck mass $M_P \sim 10^{19}$ GeV, we obtain the constraint
$$
\kappa \equiv {\Lambda_R}/{\lambda_L} < 10^3 \ .
\eqno(5.8)
$$
Since the case $\kappa \sim 1$ is experimentally ruled out,
we conclude that
$$
O(10^3)\ {\rm GeV} < \Lambda_R < O(10^5)\ {\rm GeV} \ .
\eqno(5.9)
$$
{}From (3.10), we estimate 
$$
\frac{\lambda_D}{\kappa} \simeq \frac{1}{2} \left( 1-\frac{1}{R}
\right) \frac{\lambda_R}{\kappa^2} \tan 2\theta \sim 10^9 \ .
\eqno(5.10)
$$
On the other hand, we have known that 
$$
{\frac{\Lambda_R}{\Lambda_F}}=
{\frac{\kappa}{\lambda_F}}\sim
10^{-2}
\eqno(5.11)
$$
from the study of the quark mass spectrum \cite{KF-zp}.  
Therefore, we cannot take an idea that the Dirac masses
$M_D$ and $M_F$ ($F\neq N$) are generated at the same
energy scale $\mu= \Lambda_D = \Lambda_F$.

Note that in the conventional universal seesaw model, the neutrino masses 
are  of the order of $\Lambda_L^2/\Lambda_{NR}=m_0/\lambda_L$, because of 
$M(\nu_L){\simeq}m_LM_L^{-1}m_L^T$, so that we consider $\lambda_L\sim10^9$.  
In contrast with the conventional model, in the present model, the value 
of $\lambda_L$ is $\lambda_L \sim \lambda_R/\kappa^2 \sim 10^{11}$.
Therefore, for example, the conclusion on the intermediate energy scales
based on the SO(10)$_L\times$SO(10)$_R$ model in Ref.~\cite{KoideSO10} 
is not applicable to the present model, because in Ref.~\cite{KoideSO10} 
the solutions have been investigated 
under the condition $\lambda_L\sim10^9$.  
It is a future task to seek for a unification model which satisfies 
these constraints on the intermediate 
energy scales, (5.5) and (5.7)-(5.11).

\newpage

\centerline{\Large\bf Acknowledgments}

One of the authors (Y.K.) would like to thank Professor O.~Yasuda for
his helpful comments on the cosmological constraints on the neutrino 
masses and informing the references \cite{BBN} and \cite{Fukugita}.
He also thanks Professor M.~Tanimoto and Professor A.~Yu.~Smirnov 
for pointing out a mistake (the sign of the $\Delta m^2_{solar}$) in
the first version of the paper. 
A.G. is supported by the Japan Society for Promotion
of Science (JSPS), Postdoctoral Fellowship for Foreign Researches 
in Japan  (Grant No.~99222).



\vspace{.1in}



\begin{thebibliography}{99}
%
%
\bibitem{USM} Z.~G.~Berezhiani, Phys.~Lett.~{\bf 129B}, 99 (1983);
Phys.~Lett.~{\bf 150B}, 177 (1985);
D.~Chang and R.~N.~Mohapatra, Phys.~Rev.~Lett.~{\bf 58},1600 (1987); 
A.~Davidson and K.~C.~Wali, Phys.~Rev.~Lett.~{\bf 59}, 393 (1987);
S.~Rajpoot, Mod.~Phys.~Lett. {\bf A2}, 307 (1987); 
Phys.~Lett.~{\bf 191B}, 122 (1987); Phys.~Rev.~{\bf D36}, 1479 (1987);
K.~B.~Babu and R.~N.~Mohapatra, Phys.~Rev.~Lett.~{\bf 62}, 1079 (1989); 
Phys.~Rev. {\bf D41}, 1286 (1990); 
S.~Ranfone, Phys.~Rev.~{\bf D42}, 3819 (1990); 
A.~Davidson, S.~Ranfone and K.~C.~Wali, 
Phys.~Rev.~{\bf D41}, 208 (1990); 
I.~Sogami and T.~Shinohara, Prog.~Theor.~Phys.~{\bf 66}, 1031 (1991);
Phys.~Rev. {\bf D47}, 2905 (1993); 
Z.~G.~Berezhiani and R.~Rattazzi, Phys.~Lett.~{\bf B279}, 124 (1992);
P.~Cho, Phys.~Rev. {\bf D48}, 5331 (1993); 
A.~Davidson, L.~Michel, M.~L,~Sage and  K.~C.~Wali, 
Phys.~Rev.~{\bf D49}, 1378 (1994); 
W.~A.~Ponce, A.~Zepeda and R.~G.~Lozano, 
Phys.~Rev.~{\bf D49}, 4954 (1994).
%
\bibitem{KF-zp} Y.~Koide and H.~Fusaoka, Z.~Phys. {\bf C71}, 459 (1996); 
Prog.~Theor.~Phys. {\bf 97}, 459 (1997).
\bibitem{CKM} N.~Cabibbo, Phys.~Rev.~Lett.~{\bf 10}, 531 (1996); 
M.~Kobayashi and T.~Maskawa, Prog.~Theor.~Phys.~{\bf 49}, 652 (1973).
%
\bibitem{SO10} A.~Davidson and K.~C.~Wali, in Ref.~\cite{USM};
P.~Cho, in Ref.~\cite{USM}.
Also see, Y.~Koide, Euro.~Phys.~J. {\bf C9}, 335 (1999);
Phys.~Rev. {\bf D61}, 035008 (2000).
%
%
\bibitem{Koide-nu} Y.~Koide, Phys.~Rev. {\bf D57}, 5836 (1998).
%
\bibitem{atm} Y.~Fukuda {\it et al.}, Phys.~Lett. {\bf B335}, 
237 (1994);
Super-Kamiokande collaboration, Y.~Fukuda, {\it et. al.},
Phys.~Rev.~Lett. {\bf 81}, 1562 (1998).
%
\bibitem{solar} GALLEX collaboration, P.~Anselmann {\it et al.}, 
Phys.~Lett. {\bf B327}, 377 (1994); {\bf B357}, 237 (1995); 
SAGE collaboration, J.~N.~Abdurashitov {\it et al.}, {\it ibid.} 
{\bf B328}, 234 (1994); 
Super-Kamiokande collaboration, Y.~Suzuki, in {\it Neutrino 98};
Super-Kamiokande collaboration, Y.~Fukuda {\it et al.}, 
Phys.~Rev.~Lett. {\bf 81}, 1158 (1998); {\bf 82}, 1810 (1999).
%
\bibitem{LSND} C.~Athanassopoulos {\it et al.}, Phys.~Rev.~Lett. 
{\bf 75}, 2650 (1995);  
Phys.~Rev.~Lett. {\bf 77}, 3082 (1996); nucl-ex/9706006 (1997);
G.~Mills, Talk presented at {\it Neutrino 2000},
Sudbury, Canada, June 2000 (http://nu2000.sno.laurentian.ca/).
%
\bibitem{Zee} A.~Zee, Phys.~Lett. {\bf 93B}, 389 (1980).
%
\bibitem{recentZee} For example, see 
A.~Yu.~Smirnov and M.~Tanimoto, Phys.~Rev. {\bf D55}, 1665 (1997);
C.~Jarlskog, M.~Matsuda, S.~Skadhauge, M.~Tanimoto, 
Phys.~Lett. {\bf B449}, 240 (1999);
P.~Frampton, S.~L.~Glashow, Phys.~Lett. {\bf B461}, 95 (1999)G
K.~Cheung and O.~C.~W.~Kong, Phys.~Rev.  {\bf D61}, 113012 (2000):
N.~Haba, M.~Matsuda, M.~Tanimoto  Phys.~Lett. {\bf B478}, 351 (2000); 
D.~Chang and A.~Zee, Phys.~Rev. {\bf D61}, 071303 (2000). 
%
\bibitem {Hata} N.~Hata and P.~Langacker, 
Phys.~Rev. {\bf D50}, 632 (1994); {\bf D52}, 420 (1995).
%
\bibitem{MSW} S.~P.~Mikheyev and A.~Yu.~Smirnov, 
Yad.~Fiz. {\bf 42}, 1441 (1985); 
[Sov.~J.~Nucl.~Phys. {\bf 42}, 913 (1985)]; 
Prog.~Part.~Nucl.~Phys. {\bf 23}, 41 (1989); 
L.~Wolfenstein, Phys.~Rev. {\bf D17}, 2369 (1978); {\bf D20}, 2634 (1979);
T.~K.~Kuo and J.~Pantaleon, Rev.~Mod.~Phys. {\bf 61}, 937 (1989). 
Also see, A.~Yu.~Smirnov, D.~N.~Spergel and J.~N.~Bahcall, 
Phys.~Rev. {\bf D49}, 1389 (1994).
%
\bibitem{Bahcall} J.~N.~Bahcall, P.~I.~Krastev and A.~Yu.~Smirnov,
Phys.~Rev. {\bf D58}, 096016 (1998).
%
\bibitem{atm-nu2000} H.~Sobel, Talk presented at {\it Neutrino 2000},
Sudbury, Canada, June 2000 (http://nu2000.sno.laurentian.ca/).
%
%
%
\bibitem{Karmen} K.~Eitel, Talk presented at {\it Neutrino 2000},
Sudbury, Canada, June 2000 (http://nu2000.sno.laurentian.ca/).
%
\bibitem{BBN} R.~Barbieri and A.~Dolgov, Phys.~Lett. {\bf B237},
440 (1990); 
K.~Kainulainen,  Phys.~Lett. {\bf B244}, 191 (1990); 
%
\bibitem{Fukugita} M.~Fukugita, G.~C.~Liu and N.~Sugiyama,
Phys.~Rev.~Lett. {\bf 84}, 1082 (2000).
%
\bibitem{CHOOZ} CHOOZ Collaboration, M.~Apollonio {\it et al.},
Report No. hep-ex/9907037.
%
\bibitem{KoideSO10} Y.~Koide, Euro.~Phys.~J. {\bf C9}, 335 (1999);
Phys.~Rev. {\bf D61}, 035008 (2000).
\end{thebibliography}
\end{document}